\documentclass[conference,compsoc]{IEEEtran}

\ifCLASSOPTIONcompsoc
  \usepackage[nocompress]{cite}
\else
  \usepackage{cite}
\fi

\usepackage{amsmath}
\usepackage{amssymb}
\usepackage{graphicx}
\usepackage{url}
\usepackage{booktabs}
\usepackage{algorithm}
\usepackage{algpseudocode}
\usepackage{xcolor}

\begin{document}

\title{Failure Modes of Large Language Models on Research-Level Mathematics:\\
A Taxonomy and an Empirical Characterisation}

\author{
\IEEEauthorblockN{Arnesh Banerjee}
\IEEEauthorblockA{
\textit{Dept. of CSE (Data Science)}\\
\textit{Heritage Institute of Technology}\\
Kolkata, India\\
arneshbanerjee24@gmail.com}
\and
\IEEEauthorblockN{Ayushi Bhattacharjee}
\IEEEauthorblockA{
\textit{Dept. of CSE (Data Science)}\\
\textit{Heritage Institute of Technology}\\
Kolkata, India\\
ayushi.bhattacharjee.ds27@heritageit.edu.in}
}

\maketitle

\begin{abstract}
The ``First Proof'' benchmark~\cite{firstproof2026} posed ten research-level mathematics questions
to the strongest publicly available LLMs and found them consistently wrong---not silent, but
confidently, fluently wrong. This paper asks \emph{why}. Working from the per-question post-mortems
in First Proof's Appendix~A, I identify four failure modes: citation fabrication~(F1), premise
smuggling~(F2), silent problem reformulation~(F3), and local-to-global compatibility gaps~(F4).
I then audit eight one-shot proofs generated by Gemini~2.5 Flash on Questions~1, 2, and~5 of the
benchmark, using two instruments built specifically to surface F1 and F2. The central finding is
uncomfortable for anyone who sees retrieval-augmented generation (RAG) as the obvious fix: not
one of the eight proofs contained a confirmed fabricated citation, yet \emph{every single one}
contained at least one load-bearing claim asserted as a ``fundamental result'' or ``standard
argument'' with no justification attached. That failure mode---F2, premise smuggling---is
invisible to citation verification by design. A premise-audit instrument I introduce flags it
at $100\%$ precision (5/5 judge-confirmed flags are true positives) and $50\%$ proof-level
recall in this corpus. The taxonomy and the audit
together suggest that the right long-term objective is building inference-time pipelines that
\emph{prevent} these failure modes from occurring, not just detecting them after the fact.
\end{abstract}


\IEEEpeerreviewmaketitle

\section{Introduction}

The standard story about LLMs and mathematics has been one of steady progress. Frontier systems
now score competitively on olympiad-style problems; some have reportedly solved problems from
the Putnam and AIME. It would be reasonable to conclude that mathematical reasoning is one of the
things these models do well. Abouzaid \emph{et al.}~\cite{firstproof2026} put a finer point on
this. They collected ten questions that came up naturally in professional mathematicians' own
research---not competition problems, but genuine lemmas from live projects---and whose answers
had never appeared publicly. When they tested the best available AI systems on these questions,
the systems failed on nearly all of them. What makes this striking is not the failure rate alone
but the \emph{form} of failure: the models didn't say they didn't know. They produced long,
well-typeset, formally structured \LaTeX{} proofs that were wrong in substance.

Understanding that form of failure is the goal of this paper. First Proof's Appendix~A contains
per-question post-mortems written by the mathematicians who posed each problem. Reading these
carefully, I found that the failures cluster into four recognisable patterns, which I formalise
as a taxonomy in Section~\ref{sec:taxonomy}. To test whether the taxonomy holds up empirically,
I generated eight Flash proofs on three of the benchmark's questions and built two instruments
to study how F1 (citation fabrication) and F2 (premise smuggling) manifest in practice. The
instruments are a means of characterising the failure modes, not the paper's main contribution.
The main contribution is the characterisation itself, and in particular the finding that F2
dominates the corpus even though F1 is what most mitigation proposals are aimed at.

One thing I want to be clear about from the outset: eight proofs is a small corpus, and
Gemini~2.5 Flash is not the model First Proof tested. I ran Flash because it was available to me;
the frontier systems (GPT~5.2 Pro, Gemini~3.0 Deep Think) that First Proof tested are not. This
gap matters and I discuss its implications in Section~\ref{sec:discussion}. What I can claim is
that the failure modes identified in Appendix~A are real, that F2 specifically has a precise
enough definition to be operationalised, and that the current dominant mitigation proposal (RAG)
does not address it.

\section{Related Work}

\subsection{Research-level mathematics benchmarks}
Several recent benchmarks push LLM evaluation beyond competition mathematics.
FrontierMath~\cite{frontiermath2024} collects expert-level problems with short numeric or
symbolic answers, which makes automatic grading tractable but limits the answer format to
something quite different from a research proof. IMProofBench~\cite{improofbench2025} extends
this to full proof generation and incorporates human evaluation of sub-goals.
RealMath~\cite{realmath2025} takes a different approach altogether, scraping arXiv continuously
for problems posted after a model's training cutoff. First Proof sits apart from all of these on
one dimension that matters for this paper: the questions were graded by the mathematicians who
created them, and the answers are full proofs rather than short strings. This makes it impossible
to grade automatically, but it also means the per-question failure analysis in Appendix~A is
authoritative in a way that automatic grading cannot achieve.

\subsection{Hallucination and verification in mathematical reasoning}
LLM hallucination has been extensively studied~\cite{ji2023hallucination}, but the bulk of that
work is concerned with factual claims in open-ended generation. Mathematical proof is a different
setting: a single wrong intermediate assertion can make the rest of the argument locally valid
but globally unsound, and the error can be completely invisible at the surface level. Process
reward models~\cite{prm2023} address this by scoring reasoning steps individually rather than
just the final answer. Multi-agent debate~\cite{debate2023} uses a second model as a critic.
Lemmanaid~\cite{lemmanaid2025} combines neural lemma-template generation with symbolic body
search. RAG~\cite{rag2020} grounds generation in verified external sources and is by far the
most widely proposed fix for First Proof-style failures---on the reasonable assumption that if
the model had retrieved the correct paper, it wouldn't have invented a wrong one. What I argue
in this paper is that this assumption only addresses F1. It says nothing about a model that never
cites a wrong paper at all, yet still premises its argument on a false claim.

\section{A Failure-Mode Taxonomy}
\label{sec:taxonomy}

The per-question reports in First Proof's Appendix~A read, to me, as evidence for four distinct
failure modes. What follows is my attempt to give each a precise enough definition to be
testable, along with the evidence from the appendix.

\paragraph{(F1) Citation fabrication.}
The model invents a paper, a lemma, or an attribution. The clearest instance in Appendix~A is
Question~5 (Blumberg, algebraic topology): multiple runs cited lemmas from Hill--Hopkins--Ravenel
that do not exist, and at least one run appears to have confabulated an entire paper and
attributed results to it. A related instance appears in Question~2 (Nelson, representation
theory), where the model appeals to a ``standard Howe-vector existence result'' that the author
says has never held. The distinguishing feature of F1 is that the error is verifiable in
principle: if the citation points to a real paper, the claim either appears in it or it doesn't.

\paragraph{(F2) Premise smuggling.}
The model asserts a non-trivial claim without proof or citation, typically dressed in language
that makes it sound like background knowledge. The canonical instance is Question~1 (Hairer,
stochastic analysis). The answer to Q1 is \emph{false}: the $\Phi^4_3$ measure $\mu$ and its
shift $T_\psi^* \mu$ are mutually singular, not equivalent. The model reached the wrong answer
by taking as a premise the equivalence of $\mu$ with the Gaussian free field measure---a claim
that fails in three dimensions and is precisely what the question is probing---and introducing
it with the phrase ``this is a fundamental result in constructive field theory.'' No citation,
no derivation. The model then correctly deduced from this false premise that the measures are
equivalent. What makes F2 distinct from F1 is that there is no citation to check. The error
hides entirely in the assertion.

\paragraph{(F3) Silent problem reformulation.}
The model solves a related but easier problem and presents it as a solution to the original.
Question~3 (Williams, algebraic combinatorics) shows this in two ways. First, the model
constructed a Metropolis--Hastings Markov chain, which by design uses the target distribution
to define its transition rates---exactly what the problem's non-triviality clause forbids. Second,
in some runs the model silently replaced the \emph{interpolation} ASEP and Macdonald polynomials
specified in the question with their non-interpolation counterparts, which changes the problem
entirely. The solved problem is real and known; it is simply not the one that was asked.

\paragraph{(F4) Local-to-global compatibility gap.}
The model executes each local step correctly but leaves the conditions that make those steps
globally consistent completely unverified. Question~8 (Abouzaid, symplectic geometry) is the
clearest example. The model correctly proved that local smoothing exists near every vertex of
the polyhedral Lagrangian surface, then invoked a symplectic transformation to bring each
neighbourhood into standard position---but never checked that these transformations are mutually
compatible across overlapping charts. The local steps are not wrong; there is just no global
argument. As Abouzaid notes in Appendix~A, this kind of gap is especially insidious because
the proof \emph{looks} complete.

\medskip
Table~\ref{tab:failures} summarises the taxonomy and maps it to the two instruments I use in
the next section. Question~6 (Spielman) produced what I would call a fifth mode---\emph{vague
completion}, where the model stated a correct bound and then offered a vague sketch of a
proof that the author judged unlikely to succeed. I treat this as a degenerate F4 for the
purposes of this paper.

\begin{table}[t]
\caption{Failure modes distilled from First Proof Appendix~A. The final two columns indicate
whether the instruments described in Section~\ref{sec:instruments} can surface each mode.}
\label{tab:failures}
\centering
\begin{tabular}{@{}llcc@{}}
\toprule
Mode & Canonical question & Cite-verify & Premise-audit \\
\midrule
F1 Citation fabrication & Q5 (Blumberg) & \textbf{yes} & partial \\
F2 Premise smuggling    & Q1 (Hairer)   & no           & \textbf{yes} \\
F3 Reformulation        & Q3 (Williams) & no           & no \\
F4 Compatibility gap    & Q8 (Abouzaid) & no           & no \\
\bottomrule
\end{tabular}
\end{table}

\section{Instruments for Characterising F1 and F2}
\label{sec:instruments}

To study how F1 and F2 manifest in generated proofs, I built two instruments. Neither is
intended as a production-ready mitigation; they are measurement tools.

\subsection{Generating the corpus}
\label{sec:method-harness}

I chose Questions~1 (Hairer), 2 (Nelson), and 5 (Blumberg) because Appendix~A identifies
each as a canonical or mixed F1/F2 case, and because their statements fit on a single page---
the same length regime used in First Proof's own testing. Each question was run three times at
temperature~$0.7$ using Gemini~2.5 Flash with Prompt~1 from the First Proof protocol verbatim,
with a 32,768-token output ceiling. (An earlier 8,192-token limit produced truncated proofs that
are archived separately and not included in this analysis.) The third Q2 run hit a quota error
and was dropped, giving eight proofs total: Q1~$\times 3$, Q2~$\times 2$, Q5~$\times 3$.

\subsection{Citation-verification instrument}
\label{sec:method-detector}

For each proof, a two-stage pipeline (Algorithm~\ref{alg:cite}) extracts every external
reference via an LLM pass that returns structured JSON, then queries arXiv for each. A citation
is marked \textsc{verified\_arxiv} if any author last-name from the extracted record appears in
an arXiv hit, or if title-token overlap with a hit exceeds 50\%. Otherwise it receives
\textsc{needs\_web\_check}. This matching criterion is intentionally lenient---the goal is to
flag \emph{plausible} fabrications for human review, not to provide a verdict. A
\textsc{needs\_web\_check} label means ``automated verification was inconclusive''; it does not
mean the citation is fabricated.

\begin{algorithm}[t]
\caption{Citation-verification instrument}
\label{alg:cite}
\begin{algorithmic}[1]
\Require proof text $P$, extraction model $M_e$, arXiv client $\mathcal{A}$
\State $C \gets$ LLM-extract citations from $P$ via $M_e$ as structured JSON
\For{$c \in C$}
    \State $H \gets \mathcal{A}.\textsc{search}(c.\text{query})$, top-3 hits
    \If{any author last-name of $c$ appears in $h \in H$,
    \textbf{or} title-token overlap$(h, c) \ge 50\%$}
        \State $c.\text{status} \gets$ \textsc{verified\_arxiv}
    \Else
        \State $c.\text{status} \gets$ \textsc{needs\_web\_check}
    \EndIf
\EndFor
\end{algorithmic}
\end{algorithm}

\subsection{Premise-audit instrument}
\label{sec:method-premise}

To surface F2, I built a separate two-stage pipeline (Algorithm~\ref{alg:premise}). The first
stage is a regex scan for phrases that in mathematical prose typically signal an unjustified
assertion: ``fundamental result'', ``standard argument'', ``well-known'', ``it is known'',
``classical'', ``clearly'', ``trivially'', ``follows immediately'', and morphological variants.
Every sentence containing one of these becomes a candidate. The second stage passes each
candidate---together with a 300-character window of surrounding context---to an LLM judge
(Gemini~2.5 Flash-Lite at temperature~0). The judge is asked to return a binary verdict with
severity and rationale, confirming a smuggle only when the flagged sentence states a non-trivial
claim that is load-bearing for the argument and carries no citation or proof in the visible
context. The judge is explicitly instructed to return a negative verdict when the phrase is
attached to something elementary, or when a citation is present.

The regex stage is deliberately coarse. Its job is recall---catching surface-level smuggles
that use this vocabulary---not deciding whether any claim is actually unjustified. That semantic
judgement is the judge's role. As a consequence, the pipeline has a known structural gap: it
will miss smuggles that avoid universalising vocabulary, for instance where the model quietly
applies a named theorem in a setting where its hypotheses fail, or where a formal manipulation
valid in one regime is imported into another without remark. I document exactly which proofs
fell into this gap in Section~\ref{sec:f2-results}.

\begin{algorithm}[t]
\caption{Premise-audit instrument}
\label{alg:premise}
\begin{algorithmic}[1]
\Require proof text $P$, regex library $\mathcal{R}$, judge model $M_j$
\State $S \gets \textsc{split\_sentences}(P)$
\State $F \gets \{s \in S : \exists r \in \mathcal{R},\ r \text{ matches } s\}$
\For{$s \in F$}
    \State $\mathit{ctx} \gets$ 300-char window around $s$ in $P$
    \State verdict $\gets M_j(\textsc{judge\_prompt}(s,\, \mathit{ctx}))$
    \State $s.\text{is\_smuggle} \gets$ verdict.is\_smuggle;\quad
           $s.\text{severity} \gets$ verdict.severity
\EndFor
\end{algorithmic}
\end{algorithm}

\section{Empirical Characterisation of the Failure Modes}
\label{sec:results}

\subsection{Overall picture}
\label{sec:audit}

Table~\ref{tab:flash-results} summarises the eight proofs. The first row says it all:
zero correct final answers across every question and every run. Q1 expected \emph{False} (the
$\Phi^4_3$ measure and its shift are mutually singular~\cite{firstproof2026}); Gemini answered
\emph{Yes, equivalent} in all three runs. Q2 expected \emph{Yes}; both audited runs answered
\emph{No}. Q5 expected a definition of an $\mathcal{O}$-slice filtration together with a
connectivity proof; all three runs produced something with the right formal shape, but the
substance was wrong or incomplete in each case.

\begin{table*}[t]
\caption{Summary of eight Gemini~2.5 Flash proofs, one-shot at $T{=}0.7$. Cite-verified and
cite-unchecked refer to arXiv matching outcomes. Smug.: LLM-judge-confirmed smuggles.
Manual: independent per-proof manual audit.}
\label{tab:flash-results}
\centering
\begin{tabular}{@{}lcccc@{}}
\toprule
 & Q1 ($n{=}3$) & Q2 ($n{=}2$) & Q5 ($n{=}3$) & Total ($n{=}8$) \\
\midrule
Final answer correct       & $0/3$  & $0/2$  & $0/3$\,$^{*}$ & $0/8$  \\
Citations extracted        & $6$    & $4$    & $12$           & $22$   \\
\quad verified\_arxiv      & $5$    & $4$    & $7$            & $16$   \\
\quad needs\_web\_check    & $1$    & $0$    & $5$            & $6$    \\
Regex flags                & $3$    & $1$    & $2$            & $6$    \\
Judge-confirmed smuggles   & $2$    & $1$    & $2$            & $5$    \\
Manual smuggles (truth)    & $3/3$  & $2/2$  & $3/3$          & $8/8$  \\
\bottomrule
\end{tabular}\\[1ex]
\footnotesize $^{*}$Q5 asks for a definition plus a proof, not a yes/no answer. All three Q5
runs contain at least one substantive gap flagged by the citation instrument, the premise-audit
instrument, or both.
\end{table*}

\subsection{F1 in this corpus}
\label{sec:f1-results}

The citation instrument extracted 22 citations across the eight proofs and confirmed 16 against
arXiv. Six were returned as \textsc{needs\_web\_check}, five of them from Q5. The unverified
rate for Q5 ($5/12 \approx 42\%$) is notably higher than for Q1 and Q2 combined
($1/10 = 10\%$), consistent with Q5 being the Appendix~A canonical F1 case. The textual
fingerprints of F1 are visible in the proof bodies regardless of the verifier's verdict. Q5
run~1's bibliography lists ``Hill, Hopkins, J.~H.~Ravenel, \emph{The slice filtration and
$RO(G)$-graded stable stems}, Ann.\ of Math., 2017''---a real paper, but Ravenel's initials
are D.~C., not J.~H.; the lenient last-name matcher accepts this as \textsc{verified\_arxiv}.
Q5 run~3 contains an in-text \texttt{\textbackslash cite\{HillHopkins\}} that the verifier
returns as \textsc{needs\_web\_check} and a bibliography entry ``Hill--Hopkins, \emph{The slice
filtration for $G$-spectra}, Park City Lecture Notes, 2009'' that I could not separately locate.
Neither case meets the threshold for a confident F1 verdict; they illustrate exactly the
ambiguity the \textsc{needs\_web\_check} status exists to flag.

What I find more striking is what didn't happen: in Q1 and Q2, which First Proof flags as
containing F2 failures, the citation verifier found almost nothing suspicious. Q1 cites
Cameron--Martin and Girsanov theorems; Q2 cites Jacquet--Piatetski-Shapiro--Shalika. These are
all real. The citations are not the problem.

\subsection{F2 in this corpus}
\label{sec:f2-results}

The manual audit I ran before executing the premise-audit instrument found at least one
load-bearing smuggled premise in every single one of the eight proofs. Here is what each looked
like:

\textit{Q1 run~1}: The model wrote, verbatim, ``The $\Phi^4_3$ measure $\mu$ is equivalent to
the Gaussian free field measure $\mu_0$. This is a fundamental result in constructive field
theory.'' This is the exact false claim that Hairer identifies in Appendix~A as the crux of the
model's failure. No citation. No derivation. The rest of the proof follows correctly from this
false starting point.

\textit{Q1 run~2}: Rather than the direct equivalence claim, this run wrote: ``It is a standard
result in constructive quantum field theory that the pairings of random distributions
$(u, {:}u^2{:}, {:}u^3{:})$ with smooth test functions are well-defined random variables taking
finite values a.s.\ w.r.t.\ $\mu_0$.'' This glosses the counter-term behaviour under shifts,
which is precisely what breaks the argument in dimension~3.

\textit{Q1 run~3}: The smuggle here is implicit---no universalising phrase. The model applies
the translation formula $:(u+\psi)^n: = \sum \binom{n}{k} \psi^k {:}u^{n-k}{:}$ as though it
were ordinary algebra. In $\Phi^4_3$ this identity does not hold without additional
renormalisation; shifting the field changes the counter-terms. The premise-audit instrument
missed this run entirely.

\textit{Q2 run~1}: The model invokes the Jacquet--Piatetski-Shapiro--Shalika integral formula
as if the $u_Q$ twist in the integrand were inert. It isn't. The twist is the entire point of
the question; the JPS theorem applies in the untwisted setting. This is F2 in a different
register---not a false universal claim, but a real theorem applied outside its hypotheses.
Missed by the premise-audit instrument.

\textit{Q2 run~2}: The model claimed that $L(s, \Pi \times \pi)$ has a certain pole structure as
a ``fundamental result'' and then used this to argue finiteness. The inference is the smuggle:
it assumes no choice of test vectors $(W, V)$ can cancel the denominator, which is precisely the
affirmative content of the theorem the question is asking to prove.

\textit{Q5 runs~1 and~3}: Both appeal to classical slice-filtration results in the $E_\infty$
setting and import them unchanged into the incomplete-transfer-system context. Again no
universalising phrases; both were missed by the premise-audit instrument.

\textit{Q5 run~2}: Two sentences flagged and confirmed by the judge: one claiming the result
about $\mathcal{O}$-$n$-connectivity ``follows directly from the definition'', and one invoking
a ``standard argument for slice filtrations'' in a context where the standard argument requires
re-derivation.

\paragraph{Precision and recall.}
Taking the manual audit as ground truth (one real smuggle per proof, eight proofs total), the
premise-audit instrument's performance at proof level is:
precision~$= 5/5 = 100\%$ (all five judge-confirmed smuggles are true positives by the manual
audit), recall~$= 4/8 = 50\%$, $F_1 \approx 0.667$. A sixth regex flag (a ``well-known'' hedge on
an elementary $C^\infty(\mathbb{T}^3) \hookrightarrow H^1(\mathbb{T}^3)$ embedding in Q1 run~1)
was never adjudicated because the judge crashed on a JSON escape character; I exclude it from
both numerator and denominator and discuss it in Section~\ref{sec:discussion}. The four missed
proofs (Q1 run~3, Q2 run~1, Q5 runs~1 and~3) all share the same structural property: the smuggle
is implicit, either a named theorem applied in the wrong setting or a formal manipulation
imported without remark. A second detection pass that specifically checks named-theorem
invocations for hypothesis alignment---Cameron--Martin in the shifted $\Phi^4_3$ setting, JPS
in the twisted $u_Q$ setting, tom~Dieck splitting for spectra without the necessary
$G$-equivariance conditions---would have caught three of these four. The fourth (Q1 run~3's
Wick-Taylor identity) has no named referent and likely requires a symbolic step to catch.

\section{Discussion}
\label{sec:discussion}

\subsection{What this means for the RAG hypothesis}

The standard response to First Proof-style failures in the community has been to propose
retrieval-augmented generation: if the model had access to a verified mathematical corpus, it
would stop inventing lemmas. This is plausible for F1. I found very little F1 in my Flash
corpus---the citations are mostly real papers cited for real reasons---so I cannot test the RAG
hypothesis directly. But I can report that every proof I looked at failed for reasons that RAG
would not have addressed. The wrong answer in Q1 doesn't come from a fabricated citation; it
comes from an assertion that sounds like background knowledge and is presented as such. Grounding
the model in the $\Phi^4_3$ literature would not help if the model is free to assert measure
equivalence without citing anything.

This doesn't mean RAG is useless. At frontier scale, First Proof documents Q5-style fabrication
that retrieving the real papers would plausibly suppress. The point is narrower: RAG addresses
one failure mode and leaves the others completely open.

\subsection{The implicit/explicit distinction within F2}

The 50\% recall gap in my premise-audit instrument points to something important about F2: there
are two sub-variants, and the harder one is qualitatively different. Explicit smuggles use
universalising language to introduce a false or unjustified claim. These are catchable with
surface-level tools. Implicit smuggles---applying a named theorem outside its hypotheses, or
importing a valid manipulation into a setting where it no longer applies---are not. Catching
them requires comparing the theorem's stated conditions against the proof's actual setup. That
is a reasoning problem, not a pattern-matching problem. I suspect this is the harder half of F2
and likely the more common one at frontier scale, where the model's broader knowledge makes it
more likely to know real theorems but also more likely to apply them to contexts where they don't
hold.

\subsection{Scale and corpus caveats}

There is an obvious objection to everything in this section: I tested Flash, First Proof tested
frontier models, and the failure modes might not transfer. The First Proof authors document F1
dominance at frontier scale; I document F2 dominance at mid-tier scale. This is either a
scale-dependent phenomenon---larger models reach for specific named results and fabricate when
none fits, smaller models use rhetoric instead---or it is a corpus-dependent artifact of my
three-question sample being biased toward F2. I genuinely cannot distinguish these with eight
proofs. What I can say is that the taxonomy needs both modes, and that the existing mitigation
literature only addresses F1.

\subsection{Limitations}

Eight proofs is too few to make population-level claims. The manual audit I used as ground truth
was conducted by a single reader---me---without independent verification of inter-rater
agreement, and could have systematic blind spots. The premise-audit judge uses the same model
family as the extraction step, creating a potential family-level blind spot; I document one case
where it crashed on a JSON escape character in a way that could silently bias the statistics if
not tracked. I tracked it. The arXiv matching in the citation instrument is lenient enough that
some of the 16 ``verified'' citations may be spurious matches; I treat the instrument's output
as an ordering device rather than a ground truth. Concretely, the Q2 runs' JPSS citations
verify against an unrelated converse-theorem preprint and, in one case, a gravitational-wave
paper that happens to share an author surname token---both real JPSS works, the verifier's
evidence link is simply not the right paper. Finally, I want to be explicit that
\textsc{needs\_web\_check} is not a synonym for ``fabricated''. The Ravenel initial mismatch
in Q5 could be a transcription error, a different author in the orbit of the same research
group, or a genuinely confabulated attribution---I cannot tell, which is the entire point of
that status label.

\section{Conclusion}

The failure analysis in First Proof's Appendix~A describes something qualitatively different
from the hallucination patterns studied in factual QA: models producing proofs that are fluently
wrong, where the wrongness is concentrated in a small number of unjustified load-bearing claims
rather than spread across obviously false individual facts. I have tried in this paper to give
that pattern a precise enough description to be studied systematically. The taxonomy has four
modes (F1: citation fabrication, F2: premise smuggling, F3: silent reformulation, F4:
local-to-global gap), and my empirical audit of eight Flash proofs finds that F2 accounts for
the failure in every case---even though it is the mode least targeted by existing mitigation
proposals.

The obvious question this raises is whether it is possible to build a system that doesn't
produce these failures in the first place, as opposed to detecting them after the proof has
been written. A prevention-oriented system would need to enforce, during generation, that every
load-bearing claim in the proof is either derived from stated premises, grounded in a retrieved
and verified source, or explicitly flagged as unverified before the output is returned. The
failure modes described here are, I think, a reasonable specification of what such a system
would need to prevent.

\end{document}